\documentclass[preprint,number,12pt]{elsarticle}  % review al posto di preprint,number etc.

\usepackage{amssymb}
\usepackage{amsmath,amssymb,amsfonts}
\usepackage{amsthm}
\usepackage{dsfont}
\usepackage{graphicx}
\usepackage{subfigure}
\usepackage{amsbsy}
\usepackage{mathrsfs}
\usepackage{lipsum}
\usepackage{color}
\usepackage{hyperref}

\journal{MDPI: Sensors}

\usepackage{comment}
\usepackage{siunitx}
\usepackage{lineno}

\usepackage{algorithm}
\usepackage{algorithmic}
\usepackage[margin=2.5cm]{geometry}

\newtheorem{Def}{\textbf{Definition}}
\newtheorem{Asm}{\textbf{Assumption}}
\newtheorem{Problem}{\textbf{Problem}}

\renewcommand{\imath}{\boldsymbol{\mathrm{i}}}

\begin{document}

\begin{frontmatter}

\title{Efficient Sensors Selection for Traffic Flow Monitoring: An Overview of Model-Based Techniques Leveraging Network Observability} %\tnoteref{mytitlenote}
%\tnotetext[mytitlenote]{Fully documented templates are available in the elsarticle package on \href{http://www.ctan.org/tex-archive/macros/latex/contrib/elsarticle}{CTAN}.}

%% Group authors per affiliation:
\author{Marco Fabris\fnref{myfootnote}, Riccardo Ceccato and Andrea Zanella}
%\address{Department of Information Engineering, University of Padova, via Gradenigo 6/B, Padova, 35131, Italy.}
\fntext[myfootnote]{M.~Fabris and A.~Zanella are with the Department of Information Engineering at the University of Padova (UNIPD - DEI), Padua, 35131, Italy (e-mails: marco.fabris.1@unipd.it,  andrea.zanella@unipd.it). R.~Ceccato is with the Department of Civil, Environmental and Architectural Engineering, University of Padova (UNIPD - DICEA), Via Marzolo 9, 35131 Padua, Italy (e-mail: riccardo.ceccato@unipd.it).}

%% or include affiliations in footnotes:
%\author[mymainaddress,mysecondaryaddress]{Marco Fabris}
%\ead[url]{www.elsevier.com}
%
%\author[mysecondaryaddress]{Marco Fabris\corref{mycorrespondingauthor}}
%\cortext[mycorrespondingauthor]{Corresponding author}
%\ead{support@elsevier.com}
%
%\address[mymainaddress]{1600 John F Kennedy Boulevard, Philadelphia}
%\address[mysecondaryaddress]{360 Park Avenue South, New York}
	
\begin{abstract}
The emergence of 6G-enabled Internet of Vehicles (IoV) promises to revolutionize mobility and connectivity, integrating vehicles into a mobile Internet of Things (IoT)-oriented wireless sensor network (WSN). Meanwhile, 5G technologies and mobile edge computing further support this vision by facilitating real-time connectivity and empowering massive access to the Internet. Within this context, IoT-oriented WSNs play a crucial role in intelligent transportation systems, offering affordable alternatives for traffic monitoring and management.  Efficient sensor selection thus represents a critical concern while deploying WSNs on urban networks. In this paper, we provide an overview of such a notably hard problem. The contribution is twofold: (i) surveying state-of-the-art model-based techniques for efficient sensor selection in traffic flow monitoring, emphasizing challenges of sensor placement, and (ii) advocating for
{the development of} data-driven methodologies to enhance sensor deployment efficacy and traffic modeling accuracy. Further considerations underscore the importance of data-driven approaches for adaptive transportation systems aligned with the IoV paradigm.
\end{abstract}

\begin{keyword}
%\texttt{elsarticle.cls}\sep \LaTeX\sep Elsevier \sep template
sensor selection; traffic monitoring; system observability; wireless sensor networks; urban networks; smart city; %MDPI: We changed the keyword to lowercase. Please confirm this revision.
intelligent transportation systems; internet of things; internet of vehicles; 6G
\end{keyword}

\end{frontmatter}

\section{Introduction}

{The advent of 5G technology has reshaped communication networks, improving capacity, latency, and~mobility support~\cite{alcaraz2017leading}. Beyond~enhanced connectivity, 5G enables real-time interaction among a vast number of Internet of Things (IoT) devices, fostering innovation across domains such as healthcare and transportation~\cite{Zanella-Symbiocity}. Among~its groundbreaking features, 5G aims to extend mobility support to speeds exceeding 500~km/h for ground transportation, signifying a pivotal advancement in wireless communication standards~\cite{brochure5GEU}. A~key enabler of this transformation is Mobile-Edge Computing (MEC), which decentralizes computing by bringing cloud capabilities closer to the network edge~\cite{computing2014mobile}. By~reducing latency and enabling real-time network access, MEC is particularly valuable for mobility applications, including autonomous transportation~\cite{sabella2016mobile}.}  

{Building on 5G, next-generation networks are expected to push connectivity further, paving the way for a 6G-enabled Internet of Vehicles (IoV)~\cite{osorio2022towards}. Future 6G networks will introduce disruptive technologies such as terahertz and optical communications, enhancing data rates and spectral efficiency~\cite{Toward_6G_Networks}. In~this evolving landscape, the~IoV envisions vehicles as intelligent agents equipped with sensors, computing units, and~communication interfaces, enabling seamless interaction with infrastructure, other vehicles, and~road users through vehicle-to-everything (V2X) communication. To~support real-time decision-making, AI-driven network intelligence will play a crucial role, while the increasing connectivity demands will require robust security mechanisms to mitigate cyber threats.}  

{Within intelligent transportation systems (ITSs), wireless sensor networks (WSNs) are transforming traffic monitoring and management by offering a scalable, cost-effective alternative to traditional wired sensors, which are often constrained by high installation costs and complexity~\cite{tubaishat2009wireless}. By~embedding small, wireless sensors into roadways and intersections, WSNs enable real-time traffic data collection, congestion management, and~early warning systems for safety hazards, all while reducing infrastructure costs. Their applications range from parking monitoring to adaptive traffic control, providing accurate, high-density data that support dynamic decision-making. Moreover, as~WSN technology evolves, its integration with vehicular networks holds significant potential, bridging the gap between vehicles and infrastructure to enable bidirectional traffic information exchange. This connectivity enhances situational awareness, optimizes route planning, and~fosters proactive traffic management, paving the way for smarter, more sustainable urban mobility~\cite{musa2023sustainable}.}  

{However, deploying sensors across entire road networks is often impractical due to cost constraints. To~address this, IoT-oriented WSNs leverage spatial and temporal correlations to estimate traffic conditions at unobserved locations, reducing the number of required sensors while maintaining accurate traffic monitoring. Advanced algorithms and machine learning techniques further enhance predictive capabilities, enabling authorities to anticipate congestion and dynamically adjust traffic flow~\cite{chhatpar2018machine,owais2024deep,cheng2020classifying,viti2008sensor,bae2019spatio,ben2001network}. In~addition, IoT-oriented WSNs facilitate seamless integration with emerging technologies such as connected and autonomous vehicles, enabling dynamic routing, collision avoidance, and~cooperative driving strategies~\cite{hafner2021survey,ding2019rule,chen2021novel}.}  

{Despite these advancements, sensor deployment for traffic flow monitoring remains a fundamental challenge. In~the literature, the~so-called street sensor selection problem %MDPI: Please confirm if the italics is unnecessary and can be removed. And please cheak the full text and remove all italic format.
	(also known as the traffic sensor location problem) focuses on determining the optimal number and placement of sensors for effective traffic data collection~\cite{Traffic_Sensor_Location_Problem}. Given the impracticality of equipping all network streets and intersections with sensors, finding optimal sensor locations is crucial for traffic flow estimation and congestion management.}

\subsection*{Contribution and~Outline}

{Selecting optimal locations to deploy IoT-oriented WSNs within urban road networks typically presents an arduous task.} Leveraging advanced modeling techniques and data analytics, transportation planners strive to compensate for factors %EE:Please check that intended meaning was retained
such as cost-effectiveness, coverage requirements, and~regulatory compliance. Given the complexities involved, achieving the ideal balance between sensor density, energy demand, cost considerations, and regulatory constraints remains an ongoing challenge in urban traffic management~\cite{won2020intelligent,ramirez2021sensors}.

In light of the above considerations, the~contribution of this review is~twofold:
\begin{itemize}
	\item[(i)] We survey state-of-the-art model-based techniques leveraging system observability for efficient sensor selection in traffic flow monitoring over urban road networks. Specifically, we address the challenges of sensor placement, emphasizing the importance of the above balancing factors.
	\item[(ii)] In response to the limitations of traditional traffic modeling, we advocate for {the development of new} data-driven methodologies {based on network observability} to bypass the need for explicit model design. We then highlight the potential of these approaches for enhancing sensor deployment efficacy, traffic modeling accuracy, and~real-time scheduling, thus laying the groundwork for more adaptive transportation systems aligned with the IoV paradigm.
\end{itemize}

\indent The remainder of the paper unfolds as follows. {In Section~\ref{sec:fromnetmodeltoss}, we emphasize the challenge of modeling complex transport systems and the importance of sensor locations for traffic flow monitoring.}  Section~\ref{sec:efficient_traffic_flow_monitoring} delves into the technical intricacies of sensor placement within urban landscapes, exploring observability-based metrics and optimal sensor selection strategies, which represent the core focus of our review. Then, the~discussion in Section~\ref{sec:discussion} critically evaluates traditional traffic modeling versus data-driven methodologies, highlighting the {potential} efficacy of the latter, especially considering communication technology advancements. Finally, conclusions and future directions are sketched in Section~\ref{sec:conclusions}, touching on continued investigation into novel observability-based metrics and data-driven~strategies.

\section{From Network Models to Sensor Selection for Traffic~Monitoring}\label{sec:fromnetmodeltoss}

{Traffic monitoring relies on an accurate representation of urban networks and an effective deployment of sensors to collect relevant data. The~first step in this process is the development of a suitable network model that captures the key dynamics of traffic flow. Such models, which can be formulated at different levels of granularity, provide the foundation for understanding congestion patterns and predicting traffic conditions. Once a model is available, the~challenge shifts to selecting an optimal set of sensors that ensures accurate and cost-effective monitoring of traffic states. In~this section, we first introduce modeling approaches for urban networks, followed by a discussion on the role of sensor selection strategies for effective traffic flow monitoring.}

\subsection{Modeling and Monitoring Urban~Networks}\label{sec:modeling_urban_nets}

A transport system is a complex system, made up of multiple elements linked by mutual nonlinear interactions and feedback cycles. Therefore, modeling such a system is challenging, in~particular for urban networks, where congestion significantly affects the performance of the system. 
Specifically, a~transport system consists of two main inter-related elements: travel demand and transport supply. The~former is derived from the needs of the people to move to different places, and~depends on short-term mobility choices (e.g., frequency, travel means, time, destination, and path) and long-term decisions (e.g., car ownership and public transport subscription). The~latter is composed of physical infrastructures (e.g., roads and parking spaces), services (e.g., public transport lines), regulations (e.g., circulation and parking rules), and~prices (e.g., fuel cost, tolls, and parking prices). Travel demand and transport supply are mutually influenced: travel choices depend on the level of service of the transport supply, which is affected by travel demand flows. %Based on these considerations, modeling a transport system 
Modeling a transport system generally requires the development of a supply model, representing the characteristics of the network, a~demand model, estimating the travel demand, and~an assignment model, calculating flows on each network element, by~modeling the mutual inter-dependencies between supply and demand~\cite{de2024modelling}.

% -------------------------
%A transport system is a complex system, made up of multiple elements linked by mutual nonlinear interactions and feedback cycles~\cite{cascetta2009transportation}. 
%Therefore, modeling such a system is a challenging task, in particular for urban networks, where congestion significantly affects the performance of the system. 
%Specifically, a transport system consists of two main inter-related elements: travel demand and transport supply. The former is derived from the needs of the people to move to different places, and depends on short-term mobility choices (e.g., frequency, travel means, time, destination, path) and long-term decisions (e.g., car ownership, public transport subscription). The latter is composed of physical infrastructures (e.g., roads, parking spaces), services (e.g., public transport lines), regulations (e.g. circulation and parking rules), and prices (e.g., fuel cost, tolls, parking prices). Travel demand and transport supply are mutually influenced: travel choices depend on the level of service of the transport supply, which is affected by travel demand flows. Based on these considerations, modeling a transport system implies the development of a supply model, representing the characteristics of the network, a demand model, estimating the travel demand, and an assignment model, calculating flows on each network element, by modeling the mutual inter-dependencies between supply and demand~\cite{de2024modelling}.\\
% -------------------------
In turn, an~urban network can be divided into several links, that are road segments, to~which a time and/or monetary traveling cost is associated through specific flow-cost functions. Different approaches have been developed to model a transport system; however, their main goal is to find an equilibrium of the system, i.e.,~a configuration in which demand, path, and link flows are consistent with the travel costs that they generate in the network. Assignment models can be classified according to various factors. Uncongested and congested network assignment models assume that the cost of traveling on a link is respectively independent and dependent on flows on that link. Moreover, deterministic models assume that drivers have a perfect knowledge of travel costs, thereby generating deterministic path choices; on the contrary, stochastic models explicitly model the variability in drivers' perceptions of costs, thus accounting for probabilistic travel choices. Furthermore, a~constant within-period demand and supply is assumed by static assignment models, describing a steady-state condition of the transport system; within-period dynamic assignment models allow to reproduce the time evolution of the network conditions by~explicitly modeling the effects of traffic dynamics typical of real transport systems.
%\st{In addition, traffic flows on a network can be modeled through different techniques: macroscopic, mesoscopic and microscopic traffic simulation models.}

\begin{figure}[t]
	\begin{minipage}{0.3\textwidth}
		\centering
		\includegraphics[trim={0 3mm 0 3mm},clip,width=\textwidth]{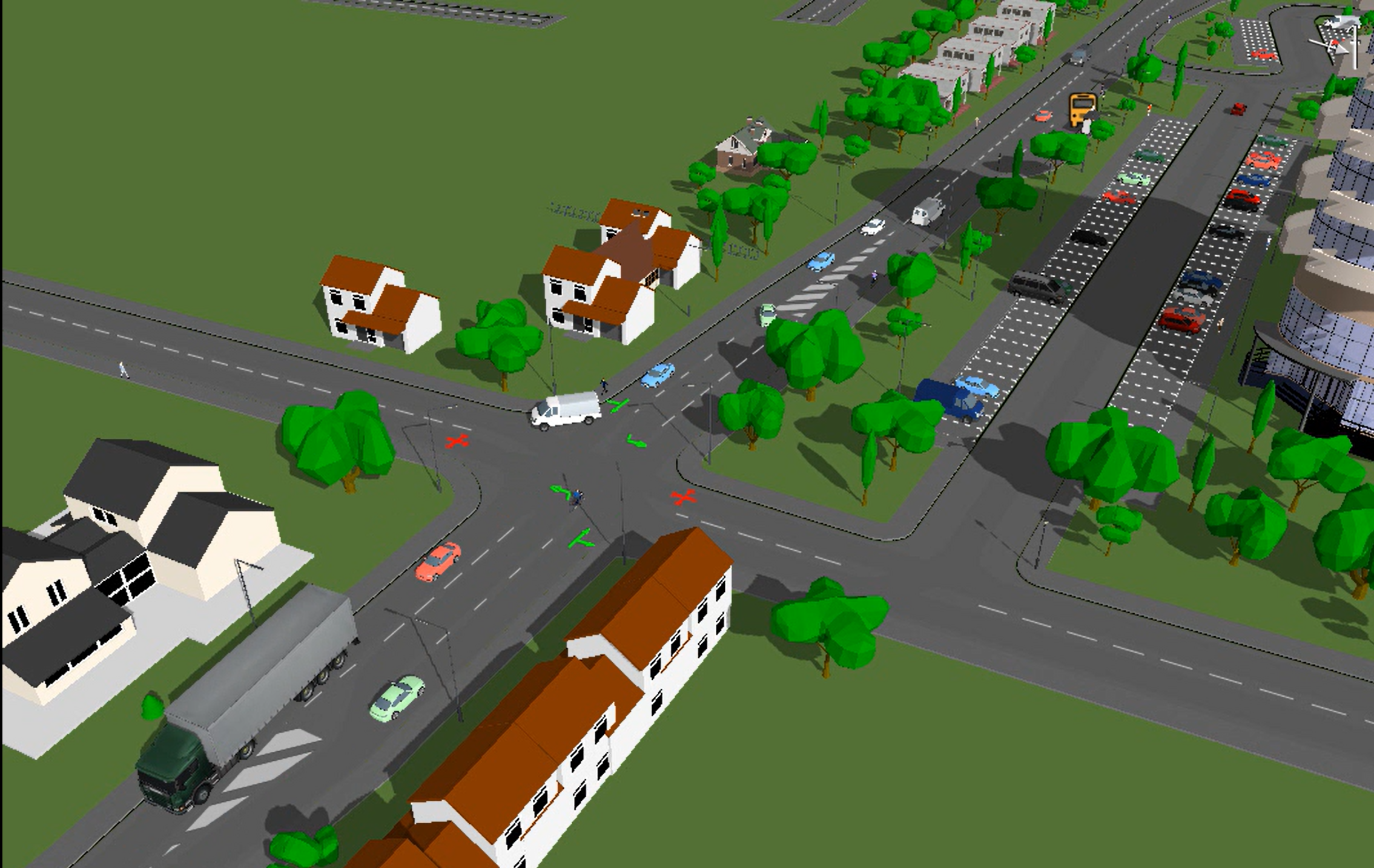}
	\end{minipage}
	\hspace{5mm}
	\begin{minipage}{0.3\textwidth}
		\centering
		\includegraphics[trim={0 0mm 0 0mm},clip,width=\textwidth]{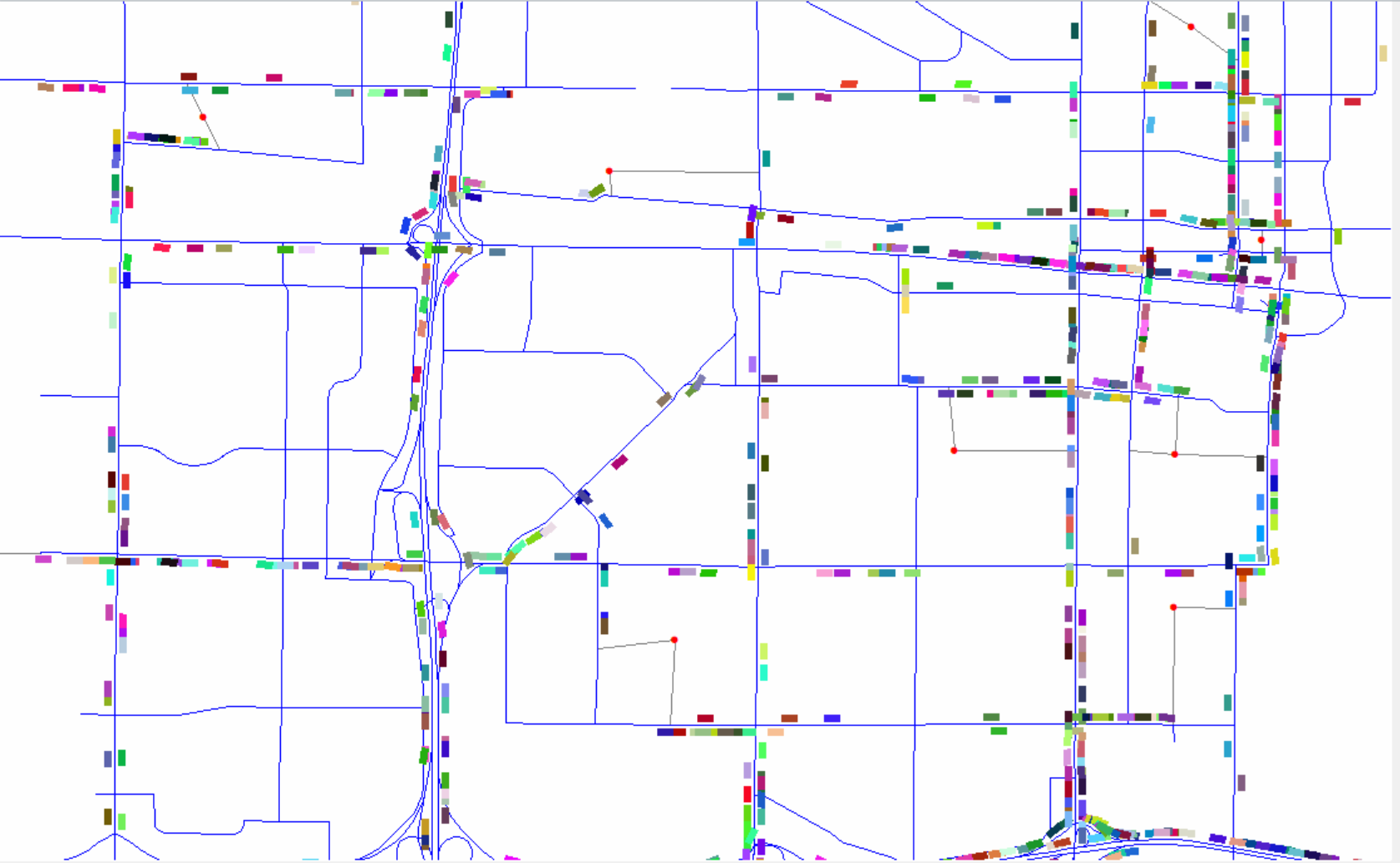}
	\end{minipage}
	\hspace{3mm}
	\begin{minipage}{0.3\textwidth}
		\centering
		% trim={<left> <lower> <right> <upper>}
		\includegraphics[trim={0 63mm 0 63mm},clip,width=0.9\textwidth]{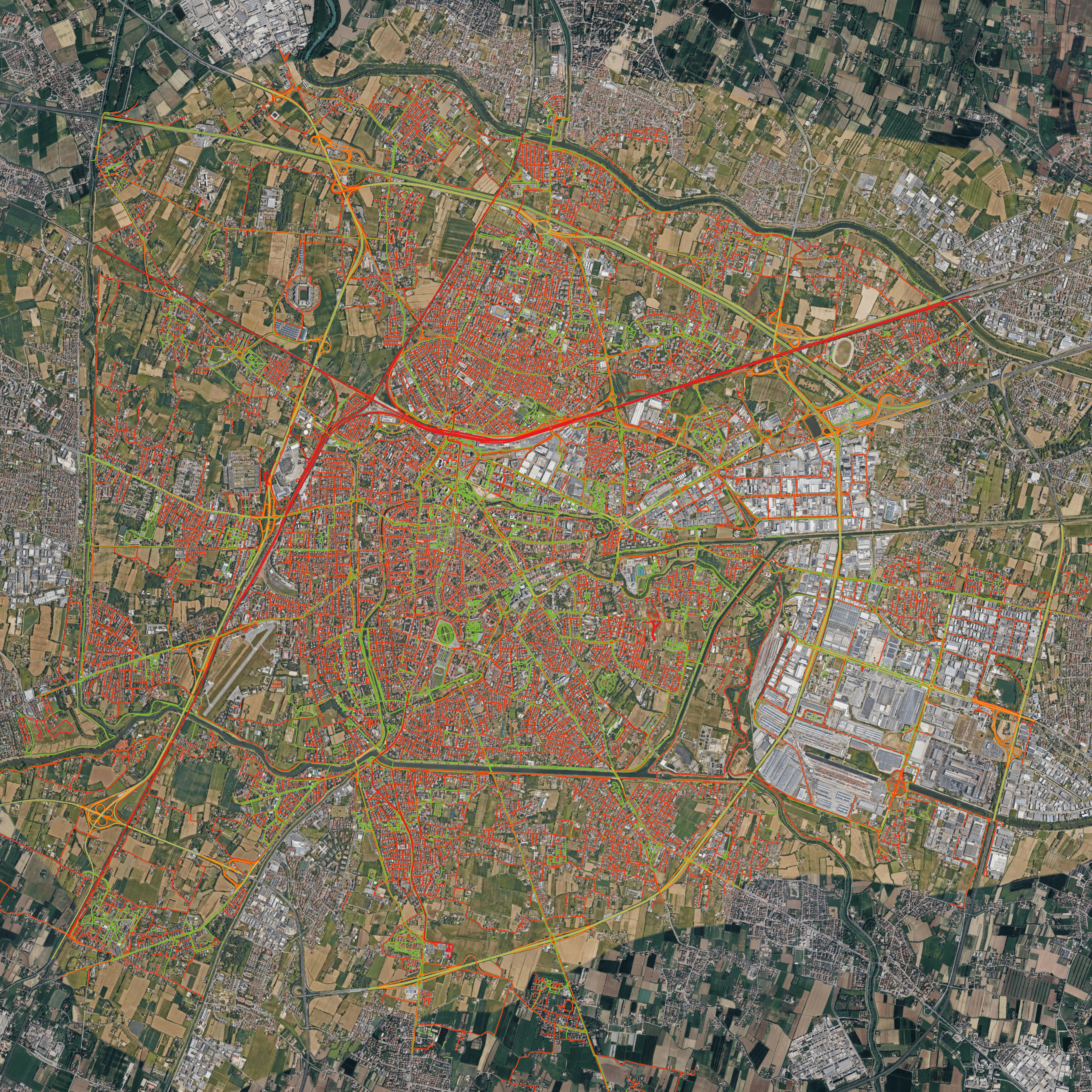}
	\end{minipage}
	\caption{From %MDPI: 1. Some contents of this figure are not legible. Please replace the image with one of a sufficiently high resolution (min. 1000 pixels width/height, or a resolution of 300 dpi or higher). 2. Please confirm if need add explanation for different colors. % No need to add explanations of different colors.
		left to right, the~representations of traffic models at different scale levels: microscale~\cite{Systra2025}, mesoscale~\cite{Bentley2020}, macroscale~\cite{Bentley2020}.}
	\label{fig:scaleleveles}
\end{figure}

Traffic flows in a network can be modeled through macroscopic, mesoscopic, and microscopic traffic simulation models~\cite{barcelo2010fundamentals} {(see Figure~\ref{fig:scaleleveles})}, whose aim is to find a configuration in which demand, path, and link flows are consistent with the travel costs that they generate in the network~\cite{cascetta2009transportation}:
\begin{itemize}
	\item Macroscopic models are based on a continuum traffic flow theory, modeling the temporal and spatial evolution of the fundamental variables describing the macroscopic flows, i.e.,~volume, speed, and density. 
	\item Microscopic models are focused on the simulation of the movement of each individual vehicle, by~modeling its actions in response to the surrounding flows, through~car-following, lane-changing, and gap-acceptance submodels. 
	\item Mesoscopic models are hybrid models; they consider a single vehicle or a group of vehicles, whose movement is based on laws describing the relationship among aggregate flow variables (e.g., average speed).
\end{itemize}

All of them share in common the travel demand expressed as an origin--destination matrix, representing the number of trips for each pair of network zones. However, the~time horizon reference of this matrix is different among the three model types: 
a unique demand matrix, constant over the entire simulation period, is used for macroscopic models, whereas a set of matrices, defined for shorter time periods, is provided for meso- and microscopic models, to~better approximate the variations in travel demand. Moreover, the~three models differ according to the network representation:
on the one hand, macro- and mesoscopic models adopt an extended and aggregated representation of nodes and links (e.g., defining capacity, number of lanes, and volume-delay function); on the other hand, microscopic models require a detailed characterization of these elements (e.g., defining lane with a speed limit, %EE:Please check that intended meaning was retained
signal settings, or specific stop rules at intersections)~\cite{barcelo2005methodological}.

\subsection{The Role of Sensor Selection in Traffic Flow~Monitoring}\label{sec:sel_prob_traffic}

The general aim of traffic sensors is to count the number of vehicles on a road section or area of the network, in~order to transform these data into traffic flows~\cite{wang2012model}. Despite this oversimplified goal, obtaining information from traffic sensors is essential to efficiently plan and manage transportation systems~\cite{fei2013vehicular}. The~accuracy and reliability of such a process are highly dependent on the quantity and quality of input data provided by monitoring sensors~\cite{ehlert2006optimisation}. However, in~practical applications, the~installation of traffic sensors in the entire network is infeasible, due to budgetary constraints~\cite{owais2019sensor}. Therefore, identifying the optimal location of a limited number of monitoring devices becomes of primary importance to obtain a trade-off between estimation accuracy of traffic states and cost associated with the implementation of a sensor system~\cite{hu2009identification}. For~these reasons, various traffic sensor location problems were proposed with different aims~\cite{Traffic_Sensor_Location_Problem}.

Forecasting the impacts of changes in transportation networks and travel demand policy is the main aim of traffic simulations~\cite{castillo2008observability}. As~described in Section~\ref{sec:modeling_urban_nets}, one of the essential inputs of such simulation models is an origin--destination matrix. A~direct estimation of this matrix based on on-site surveys (such as number plate recognition, road-side surveys, or household interviews) is often very expensive in terms of costs and time effort~\cite{cipriani2006heuristic}. For~this reason, many previous authors have focused on the optimization of the location of traffic monitoring sensors to obtain the most accurate and reliable estimate of origin--destination matrices~\cite{sun2021bi}.

Moreover, traffic flows on the entire network can be determined based on the recorded volumes of vehicles, provided by sensors installed in a limited number of locations, leading to a flow observability problem and a link flow inference problem~\cite{Traffic_Sensor_Location_Problem}. The~solution to the first expands the information on a subset of roads, through the application of flow propagation rules and assuming a predefined traveler route choice behavior~\cite{hu2014generalized}. On the other hand, the~solution to the second helps infer link flows, without~the need for prior information about route choice proportions~\cite{gentili2011survey}.  Furthermore, the~combination of data from multiple traffic sensors can be adopted to estimate network performances. In~particular, previous authors have developed sensor location problems to derive travel time on roads, considering the dynamic nature of traffic~\cite{park2015optimal}. In~addition, information from fixed sensors can be used to infer congestion levels on the network~\cite{cao2024tracking}.

Further applications of the traffic sensor location problem include path reconstruction, aiming at understanding how observed flows from a monitoring system can be used to identify the paths among different origin--destination pairs~\cite{fu2016heterogeneous}. Moreover, recorded volumes of vehicles approaching a road intersection can be considered as input of a traffic light system, where traffic light phases and operations are optimized to minimize congestion and users' delay~\cite{cruz2018optimized}.

For these reasons, defining the optimal location of sensors monitoring traffic volumes is essential to evaluate the performance of the system, both at the local and network level. In~particular, monitoring sensors can be adopted to analyze the current conditions of the road where they are installed, by~detecting vehicle queues or travel time in a corridor~\mbox{\cite{viti2008sensor,bae2019spatio}}. In~addition, sensors can be used to assess the conditions of the entire network~\cite{varotto2022VSNs} by~applying traffic simulation models~\cite{fei2011structural} or specific techniques inferring network states from local data provided by monitoring devices~\cite{Traffic_Sensor_Location_Problem}.

%\indent Once a network model is available, traffic flow monitoring can be then carried out. This is essential to obtain traffic data, that are necessary for refining the underlying network model or carry out surveillance tasks with the support of visual sensor networks~\cite{varotto2022VSNs}. In particular, traffic volumes from a set of sensors located in a road network can be used to estimate the origin-destination matrices and inputs for the traffic models~\cite{fei2011structural}, as discussed in the sequel. In addition, these data allow to directly evaluate network conditions, such as travel time, or to detect traffic queues~\cite{viti2008sensor,bae2019spatio}. 

%%%%%%%%%%%%%%%%%%%%%%%
%%%%%%%%%%%%%%%%%%%%%%%
%%%%%%%%%%%%%%%%%%%%%%%
%%%%%%%%%%%%%%%%%%%%%%%
\section{Overview of the Street Sensor Selection~Problem} \label{sec:efficient_traffic_flow_monitoring}

In this section, we explore techniques and challenges related to sensor selection within a WSN for traffic monitoring over road networks. The~solutions to this class of problems aim to identify the most suitable sensors from a pool of available options in order to perform efficient monitoring. To~this aim, we shall review the notion of observability, examine a number of observability-based metrics, and discuss the major advances obtained so far in the attempt to solve similar yet diverse versions of the street sensor selection problem.
%\AZa{Visto l'audience della conferenza, dovresti ricordare  cosa si intende formalmente per osservabilità di un sistema. }

\subsection{Mathematical~Preliminaries}\label{sec:preliminaries}

{Section~\ref{sec:sel_prob_traffic} can be summarized as follows: once a network model is available, traffic flow monitoring through sensor placement can be then carried out. Since urban networks can be described, for~instance, through state-space models, it is useful to recall some fundamental concepts of System Theory (see, e.g.,~\cite{hespanha2018linear}, for~a reference on the topic) related to linear time-invariant (LTI) systems %EE:Please check that intended meaning was retained
	and their representation. In~addition, the~concept of submodularity is defined subsequently.}

%Here we introduce basic notions of System Theory (see, e.g.~\cite{hespanha2018linear}, for a reference on the topic) related to Linear Time-Invariant (LTI) systems and their state-space representation. Also, the concept of submodularity is defined.

\subsubsection{LTI Systems and State-Space~Representations}

A system $\Sigma$ is a function whose  domain and codomain are sets of signals, %EE: Please check that intended meaning was retained
i.e.,
\begin{align}
	\Sigma:\quad & \mathcal{U} \rightarrow \mathcal{Y}, \nonumber\\
	& u(\cdot) \mapsto y(\cdot) = \Sigma (u(\cdot)),
\end{align}
where $\mathcal{U} \subseteq \mathbb{R}^m$ and $\mathcal{Y}\subseteq \mathbb{R}^p$ are said to be the input set and output set, respectively.
A system $\Sigma$ is said to be linear if, for any couple of signals $\nu_1(\cdot)$ and $\nu_2(\cdot)$ and for all $\alpha_1, \alpha_2\in \mathbb{C}$, the following property hold{s}:
\begin{equation}
	\Sigma(\alpha_1 \nu_1(\cdot) + \alpha_2 \nu_2(\cdot)) = \alpha_1 \Sigma(\nu_1(\cdot)) + \alpha_2 \Sigma(\nu_2(\cdot)).
\end{equation}
{Denoting %MDPI: Please check through the paper if indentation should be added to the first line after equations.
	with $\Delta_{t_0}\!: \nu(t) \mapsto \nu(t-t_0)$ the time-shift operator}, a~system is said to be time-invariant if, for any $\nu(\cdot)$, one has
\begin{equation}
	\Sigma(\Delta_{t_0}(\nu(\cdot))) =  \Delta_{t_0}(\Sigma(\nu(\cdot))).
\end{equation}

It is well known that convolutional linear time-invariant (LTI) systems (for %MDPI: Footnote is not permitted in this journal. We included this paragraph in the maintext. Please check the full text and confirm this revision.
a convolutional system, the output is obtained as the convolution operation between its impulse response and a given input (see, for further details,~\cite{oppenheim2017signals})) admit a description consisting of a system of first-order difference equations, called state space. By~defining the $n$-dimensional signal $x(\cdot)$, named the state, this type of representation captures the system's dynamics, which can be expressed in the following standard form (we assume the time domain to be discrete throughout this paper):
\begin{equation}\label{eq:state-space}
	\Sigma:\quad \begin{cases}
		x(t+1) = Ax(t) + Bu(t), \quad x(0) =  x_0\\
		y(t) = Cx(t) + D u(t).
	\end{cases}
\end{equation}
In particular, $x(t) \in \mathbb{R}^n$ denotes the state vector describing the system's internal state $x(\cdot)$ with a given initial condition $x_0$, while $u(t) \in \mathbb{R}^m$ and $y(t) \in \mathbb{R}^p$ denote the input and output vectors, respectively.
More specifically, the~system's dynamics are governed by the constant matrices $A \in \mathbb{R}^{n \times n}$, $B \in \mathbb{R}^{n \times m}$, $C \in \mathbb{R}^{p \times n}$, and $D \in \mathbb{R}^{p \times m}$: matrix $A$ represents the relationship between the current state and the future state of the system, while $B$ describes how the input $u(t)$ influences the state $x(t)$, matrix $C$ links the system's state to the observable output $y(t)$, and,~finally, $D$ describes the direct effect of the input, if~present, on~the output. A~common shorthand notation for the system in~\eqref{eq:state-space} is therefore given by the ordered tuple $\Sigma=(A,B,C,D)$.

The natural response $y_n(\cdot)$ and the forced response $y_f(\cdot)$ of a system $\Sigma$ can be obtained in the first place by calculating the general solution of the system described by equation \eqref{eq:state-space}, which is expressed as
\begin{equation}\label{eq:solution_ss}
	x(t) =  \underset{=:x_n(t)}{\underbrace{A^t x(0)}} + \underset{=:x_f(t)}{\underbrace{\sum_{k=0}^{t-1}A^{t-1-k}Bu(k)}}.
\end{equation}
The natural evolution of the state is thus yielded by the term $x_n(t)$, while the remaining terms form the forced evolution of the state $x_f(t)$. Consequently, in~accordance with \eqref{eq:solution_ss}, the~natural response $y_n(t)$ of an LTI system is defined as
\begin{equation} \label{eq:nat_resp}
	y_n(t) := Cx_n(t) = CA^t x(0).
\end{equation}
The signal \eqref{eq:nat_resp} describes the system's output when there are no active inputs and it is determined solely by the system's initial conditions and the internal dynamics defined by the system's state equations.
On the other hand, the~forced response $y_f(t)$ of the LTI system describes the system's output when inputs act upon it, reflecting the system's reaction to nonzero inputs without being influenced by initial conditions. Hence, for~an LTI system, the~forced response is defined as
\begin{equation}\label{eq:forced_resp}
	y_f(t) := C x_f(t) + Du(t) = C\sum_{k=0}^{t-1} A^{t-1-k}Bu(k) + Du(t),
\end{equation}
so that the relation
\begin{equation}\label{eq:output_resp}
	y(t) = y_n(t) + y_f(t)
\end{equation}
holds for all $t \in \mathbb{N}$.

\subsubsection{Set~Functions} \label{ssec:submodular}

Sensor placement problems can be formulated as set function optimization problems. Given the finite set of potential sensor locations \mbox{$\mathcal{S} = \{1,\dots,p\}$}, a~set function $f:2^\mathcal{S} \rightarrow \mathbb{R}$ assigns a real number to each subset of $\mathcal{S}$. Then, a~set function optimization problem can be formulated as
\begin{equation}\label{eq: generalOptProblem}
	\mathcal{Q}^\star = \underset{\mathcal{Q} \subseteq \mathcal{S},\quad |\mathcal{Q}|=p^\star}{\arg\max} f(\mathcal{Q})
\end{equation}
The aim is to select a $p^\star$-dimensional subset $\mathcal{Q}$ of $\mathcal{S}$ that maximizes $f$. %\footnote{In case $f$ need to be minimized, $\arg\max$ is replaced by $\arg\min$ in~(\ref{eq: generalOptProblem}).} 
In general, this kind of combinatorial optimization problem is intractable in the case of very large set cardinality $|\mathcal{S}|$; nonetheless, if~the objective function $f$ is monotone, optimization in \eqref{eq: generalOptProblem} admits a greedy algorithm to find a solution. In~particular, if~$f$ is also submodular, such a solution is achieved suboptimality within a \mbox{guaranteed bound~\cite{Intractability,summers2015submodularity}}. 
The notions of monotonicity and submodularity, along with that of modularity, are defined in the following~lines.

\begin{Def}[Monotone set function]\label{def:monotone}
	A set function $f:2^\mathcal{S} \rightarrow \mathbb{R}$ is monotone if, for all $\mathcal{A} \subseteq \mathcal{B} \subseteq \mathcal{S}$, it holds that
	\begin{equation}
		f(\mathcal{A}) \leq f(\mathcal{B}).
	\end{equation}
\end{Def}
It is evident that monotone functions exhibit growth %EE:Please check that intended meaning was retained
(in a broad sense) as any element $s\in \mathcal{S} \setminus \mathcal{A}$ is added to the current set $\mathcal{A}$.
\begin{Def}[\textbf{(Sub)modular set function}]\label{def:submodularity}
	A set function $f:2^\mathcal{S} \rightarrow \mathbb{R}$ is called submodular if, for~all subsets $\mathcal{A} \subseteq \mathcal{B} \subseteq \mathcal{S}$ and elements $s\in \mathcal{S} \setminus \mathcal{B}$, it holds that
	\begin{equation}\label{eq:submodularitydef}
		f(\mathcal{A} \cup \{s\}) - f(\mathcal{A}) \geq f(\mathcal{B} \cup \{s\}) - f(\mathcal{B}).
	\end{equation}
	If the equality in~\eqref{eq:submodularitydef} holds strictly for all subsets $\mathcal{A} \subseteq \mathcal{B} \subseteq \mathcal{S}$ and elements $s\in \mathcal{S} \setminus \mathcal{B}$, then $f$ is said~to be modular.
\end{Def}
Intuitively, the~marginal value $(f(\mathcal{A} \cup \{s\}) - f(\mathcal{A}))$ of a submodular function increases more with the addition of a new set member when the current size of the set $\mathcal{A}$ is smaller (e.g., the~function counting the number of covered subsets of a given space is monotone and submodular). On the other hand, for~a modular function, the~marginal value $(f(\mathcal{A} \cup \{s\}) - f(\mathcal{A}))$ of adding an element to set $\mathcal{A}$ is always the same, regardless of what is already in the set (e.g., the~cardinality of a set is a monotone modular function). Clearly, all modular functions are also submodular.
Lastly, it is crucial to note that submodular functions are not monotone in general. In~this case, to~maximize nonmonotone submodular functions, ad hoc heuristics or special techniques can be used (see, e.g.,~\cite{feige2011maximizing} for an in-depth discussion on that topic).

\subsection{Observability-Based Metrics and Derivation of the~Problem}\label{ssec:obs_metrics} 
In this investigation, we restrict our attention to the class of discrete-time LTI systems whose dynamics are usually described by the state-space models of the form \eqref{eq:state-space}.
Recall that the dynamics of \eqref{eq:state-space} are governed by the constant matrices $A \in \mathbb{R}^{n \times n}$, $B \in \mathbb{R}^{n \times m}$, $C \in \mathbb{R}^{p \times n}$, and $D \in \mathbb{R}^{p \times m}$ (extensions of this theoretical framework are, of course, possible, e.g.,~by considering time-varying and/or state-input-dependent matrices), given the initial condition $x(0) = x_0$.

One of the most important properties of dynamic systems is observability, i.e.,~the possibility of reconstructing or estimating $x(0)$ given the past input--output observations:
\begin{equation} \{(u(\tau),y(\tau)):~\tau=0,\ldots,t-1\}.
\end{equation} 
It is well known that observability can be proven to hold by checking the rank of the so-called observability matrix:
\begin{equation} \label{eq:observability_matrix}
	\mathcal{O} := \begin{bmatrix}
		C \\ CA \\ \vdots \\ CA^{n-1}
	\end{bmatrix}.
\end{equation}
In particular, the~system in \eqref{eq:state-space} is observable if and only if
\begin{equation}
	\mathrm{rank}[\mathcal{O}] = n.	
\end{equation}

{\subsubsection*{Example} %MDPI: Section headings should be numbered sequentially, we changed it to 3rd level section heading. Please confirm this revision. if not, please try to use a list format.
	Consider the state matrix
	\begin{equation}
		A = \begin{bmatrix}
			-0.5 & 0.25 & 0.2 \\
			0 & -0.9 & 0.1 \\
			0 & -0.9 & 0.1
		\end{bmatrix}
	\end{equation}
	corresponding to a compartmental system and set
	\begin{equation}
		C_1 = \begin{bmatrix}
			1 & 0 & 0 \\
			0 & 0 & 0 \\
			0 & 0 & 0
		\end{bmatrix}, \quad\quad 
		C_2 = \begin{bmatrix}
			0 & 0 & 0 \\
			0 & 1 & 0 \\
			0 & 0 & 1
		\end{bmatrix}.
	\end{equation}
	From the couples $(A,C_i)$, the~observability matrices $\mathcal{O}_i$ are defined respectively for $i=1,2$ according to \eqref{eq:observability_matrix}. It can be verified that $\mathrm{rank}[\mathcal{O}_1] = 3$ and $\mathrm{rank}[\mathcal{O}_2] = 2$, even though $C_2$ selects six rows over the powers of $A$ and $C_1$ selects just three rows in total. This is due to the fact that the minor
	\begin{equation}
		A_{-1} = \begin{bmatrix}
			-0.9 & 0.1 \\
			-0.9 & 0.1
		\end{bmatrix}
	\end{equation}
	represents a maximal invariant subsystem, which behaves as a ``trapped compartment''.}

{In general, if} observability is shown to hold, {a} dynamic estimate (static %MDPI: Footnote is not permitted in this journal. We included this paragraph in the maintext. Please confirm this revision.
estimates can be also obtained, but~they are either available only after a certain time delay or, when computed in an open-loop fashion, they become unreliable/ computationally demanding as $t$ grows) $\hat{x}(t)$ for the state of \eqref{eq:state-space} as $t$ varies can be computed via the Luenberger estimator~\cite{luenberger1964observing}
\begin{equation}\label{eq:state_estimator}
	\hat{\Sigma}:~\begin{cases}
		\hat{y}(t) = C\hat{x}(t) + Du(t), \quad \hat{x}(0) = \hat{x}_0 \\
		\hat{x}(t+1) = A \hat{x}(t) + B u(t) + L(\hat{y}(t)-y(t)),
	\end{cases}
\end{equation}
where $\hat{x}_0$ is the initial state estimate (this %MDPI: Footnote is not permitted in this journal. We included this paragraph in the maintext. Please confirm this revision.
value is set by either exploiting the presently available information on the system or relying on a sensible guess) and $L \in \mathbb{R}^{n \times p}$ is a suitably chosen Luenberger observer gain. Specifically, the~matrix $L$ is selected so that the estimation error
\begin{equation}
	e(t) = \hat{x}(t) - x(t)
\end{equation}
vanishes as $t$ grows, namely, the following dynamics are asymptotically stable:
\begin{equation}\label{eq:error_dynamics}
	e(t) = (A+LC)^t e(0).
\end{equation}
Denoting with $\rho(\cdot)$ the spectral radius of a matrix (i.e., the maximum eigenvalue in modulus of the spectrum), it can be proven that the error dynamics \eqref{eq:error_dynamics} are asymptotically stable if and only if $L$ satisfies
\begin{equation}
	\rho(A+LC) < 1.	
\end{equation} 
More precisely, given the system $\Sigma = (A,B,C,D)$, there exists a state estimator \eqref{eq:state_estimator} such~that
\begin{equation}
	\lim_{t \rightarrow \infty} e(t) = 0
\end{equation}
if and only if $\Sigma$ is detectable, namely $\Sigma$ is either observable or the unobservable subsystem of $\Sigma$ (the %MDPI: Footnote is not permitted in this journal. We included this paragraph in the maintext. Please confirm this revision.
unobservable subsystem of $\Sigma$ can be found by transforming its system matrices through a change of basis given by the null space of the observability matrix obtained from $(A,C)$ and its complement) is asymptotically~stable.

Observability is then closely related to state estimation performance through certain metrics computed on the observability Gramian of order $t\in \mathbb{N}\setminus \{0\}$, defined as
\begin{equation}\label{eq:obs_Gramian}
	\mathcal{W}_t := \sum\limits_{k=0}^{t-1} (A^\top)^k C^\top C A^k, \quad t=1,2,\ldots  ~. 
\end{equation}
Indeed, it is immediately verifiable %EE: Please check that intended meaning was retained
that $\mathcal{W}_n = \mathcal{O}^{\top}\mathcal{O}$. It is also worth recalling that, if~the matrix $A$ is asymptotically stable, the~infinite horizon observability Gramian
\begin{equation}
	\mathcal{W}_{\infty}:= \underset{t\rightarrow \infty}{\lim} \mathcal{W}_{t}
\end{equation}
can be computed by solving the matrix Lyapunov equation
\begin{equation}
	A{^{\top}} \mathcal{W}_{\infty} A - \mathcal{W}_{\infty} + C^{\top} C = 0.
\end{equation}

The interpretation of the observability Gramian \eqref{eq:obs_Gramian} revolves around the quadratic~form
\begin{equation}
	E(x_0,t) = x_0^\top \mathcal{W}_t x_0,
\end{equation}
which can be associated with the energy of the natural response induced by the initial state $x_0$: the larger this quantity, the~more observable the initial state is. Practically, cheaper sensing is required as an observability-Gramian-based metric indicates that the energy $~E(x_0,t)~$ increases independently from the initial state $x_0$.
A few examples of such metrics are taken from~\cite{baggio2022energy,varotto2019street,summers2014optimal,Manohar2022Optimal}, reported in the following list, and characterized in detail in Table~\ref{tab:gramian_metrics}:
\begin{itemize}
	\item $\mathrm{rank}[\mathcal{W}_n]$: the rank of $\mathcal{W}_n$ quantifies the dimension of the observable subspace. 
	\item $\mathrm{trace}[\mathcal{W}_n]/n$: the trace of $\mathcal{W}_n$ scaled by the state dimension $n$ is directly related to the average output energy and can be interpreted as the average observability in all directions of the state.
	\item $\mathcal{K}[\mathcal{W}_n]$: the condition number of $\mathcal{W}_n$ (the condition number of a matrix is defined as the ratio between the maximum and minimum singular values of a given matrix) measures how balanced the observability is among all modes. It grows unbounded if the system is unobservable. 
	\item $\lambda_{min}[\mathcal{W}_n]$: the minimum eigenvalue of $\mathcal{W}_n$ is related to the hardest state components to observe; $\lambda_{min}[\mathcal{W}_n]$ is zero for unobservable systems. Differently from $\mathcal{K}[\mathcal{W}_n]$, this metric enhances the robustness to sensor failure rather than system mode balancing.
	\item $\det[\mathcal{W}_n]^{1/n}$: whenever $\mathcal{W}_n$ is invertible, the~determinant of the observability Gramian is proportional to the volume of the ellipsoid containing the initial states that can be observed with one unit %EE: Please check that intended meaning was retained
	or less estimation energy.
	\item $\mathcal{H}_2[\Sigma]:=H_2[\mathcal{W}_{\infty}]$: whenever $D=0$, the~$\mathcal{H}_2$-norm of an LTI system is bounded. It can be proven that $H_2[\mathcal{W}_{\infty}] = \mathrm{trace}[B^\top \mathcal{W}_{\infty} B]$. The~intuition is that more potent sensors usually output stronger signals, and~this potency is captured by the $\mathcal{H}_2$-norm.
	\item $\ell d[\Sigma]:=\log\det (B^\top \mathcal{W}_{\infty} B)$: this can be intended as a volumetric variant of $\mathcal{H}_2[\Sigma]$, as~it computes the logarithm of the geometric mean of the axes of the observability Gramian ellipsoid skewed by $B$. 
\end{itemize}
\begin{table*}[t!]
	\centering
	\renewcommand{\arraystretch}{1.5}
	\begin{tabular}{|p{1.5cm}|p{5.5cm}|c|c|c|}
		\hline
		%\rowcolor[HTML]{C0C0C0} 
		\textbf{Metric}          & \textbf{Eigenvalue-based $\quad\quad\quad\quad$ characterization}                                                       & \textbf{Monotone} & \textbf{Submodular} & \textbf{Modular} \\ \hline
		$\quad\quad\quad\quad$ \(\mathrm{rank}[\mathcal{W}_n]\) & $\quad\quad\quad\quad$ $\left\|\begin{bmatrix}
			\lambda_{1}[\mathcal{W}_n]  & \cdots & \lambda_{n}[\mathcal{W}_n]\end{bmatrix}\right\|_{0}$                                               &Yes           & Yes                  & No                \\ \hline
		\(\mathrm{tr}_{n}[\mathcal{W}_n]\) & $n^{-1}\sum\nolimits_{i=1}^{n} \lambda_{i}[\mathcal{W}_n]$         &Yes    & Yes                  & Yes               \\ \hline
		$\quad\quad\quad\quad$ \(\mathcal{K}^{-1}[\mathcal{W}_n]\)     & $\quad\quad\quad\quad$ $\quad\quad\quad\quad$ $\lambda_{n}[\mathcal{W}_{n}]/\lambda_{1}[\mathcal{W}_{n}]$         &No  & No                   & No                \\ \hline
		$\quad\quad\quad\quad$ \(\lambda_{min}[\mathcal{W}_n]\) & $\quad\quad\quad\quad$ $\quad\quad\quad\quad$ $\quad\quad\quad\quad$ $\lambda_{n}[\mathcal{W}_{n}]$  &Yes & No                  & No                \\ \hline
		$\quad\quad\quad\quad$ \(\mathrm{det}_n[\mathcal{W}_n]\) & $\quad\quad\quad\quad$ $\quad\quad\quad\quad$ $\quad\quad\quad\quad$ $(\prod\nolimits_{i=1}^{n} \lambda_{i}[\mathcal{W}_n])^{1/n}$ & Yes        & Yes                  & No                \\ \hline
		\(H_2[\mathcal{W}_\infty]\) & $\sum\nolimits_{i=1}^{n} \lambda_{i}[B^{\top}\mathcal{W}_{\infty} B+D^{\top}D]$             &Yes  & Yes                  & Yes               \\ \hline
		$\quad\quad\quad\quad$ $\quad\quad\quad\quad$ \(\ell d[\mathcal{W}_\infty]\) & $\quad\quad\quad\quad\quad\quad\quad\quad\quad\quad$ $\sum\nolimits_{i=1}^{n} \log(\lambda_{i}[B^{\top}\mathcal{W}_{\infty} B +D^{\top}D ])$                         &Yes      & Yes                  & No                \\ \hline
	\end{tabular}
	\caption{Properties of metrics computed from the observability Gramian. A characterization w.r.t. to eigenvalues $\lambda_{i}[\cdot]$ of the observability Gramian or its transformations, such that $\lambda_{1}[\cdot] \geq \lambda_{2}[\cdot] \geq \cdots \geq  \lambda_{n}[\cdot]$, is also included in the second column. The third column displays whether a metric is monotonically increasing w.r.t. such a characterization and monotone according to Definition~\ref{def:monotone}. The last two columns indicate whether a metric is submodular and/or modular according to Definition~\ref{def:submodularity}.}
	\label{tab:gramian_metrics}
\end{table*}

\begin{figure}[t]
	\begin{minipage}{0.45\textwidth}
		\centering
		\includegraphics[width=\textwidth]{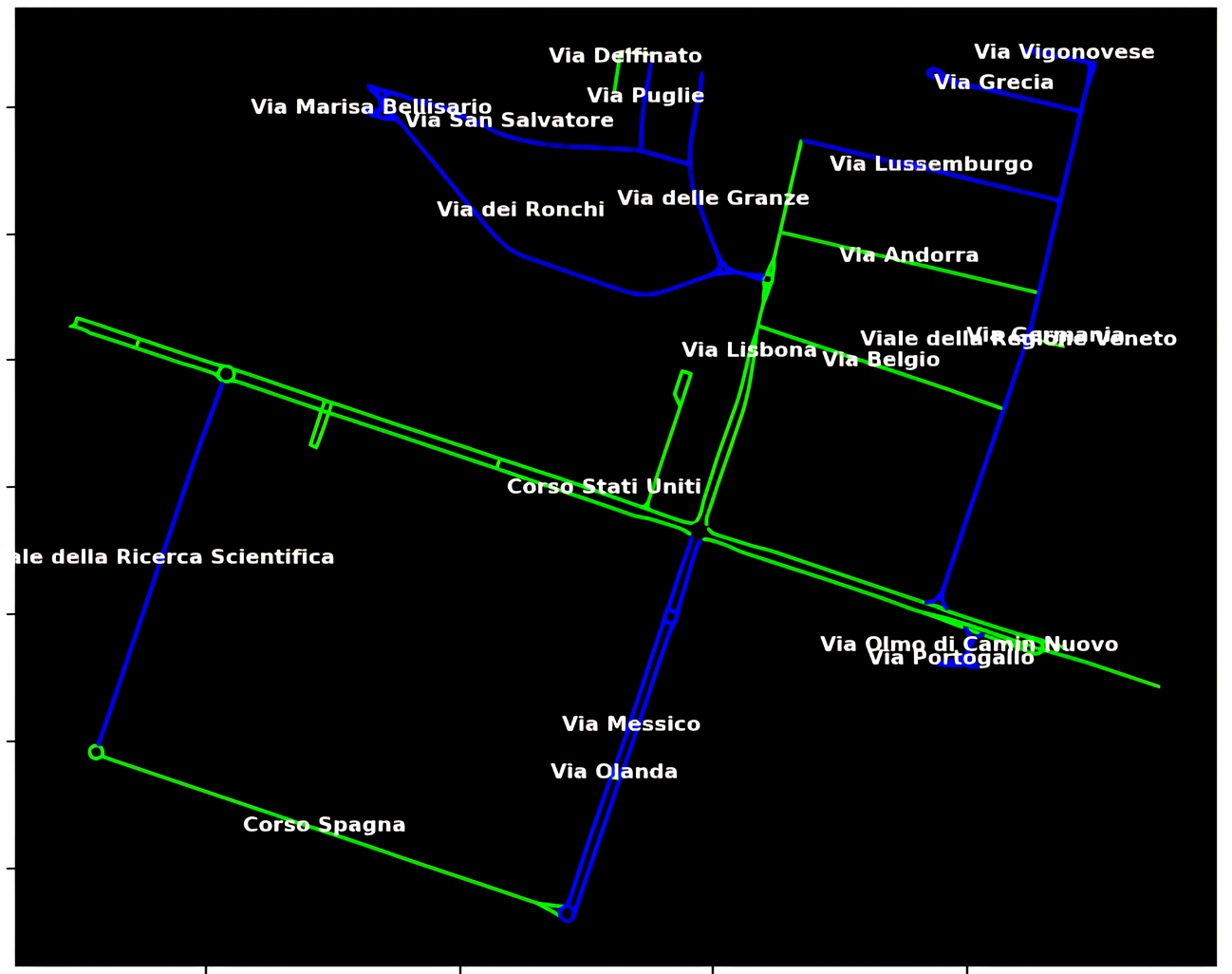}
	\end{minipage}
	\hspace{3mm}
	\begin{minipage}{0.453\textwidth}
		\centering
		\includegraphics[width=\textwidth]{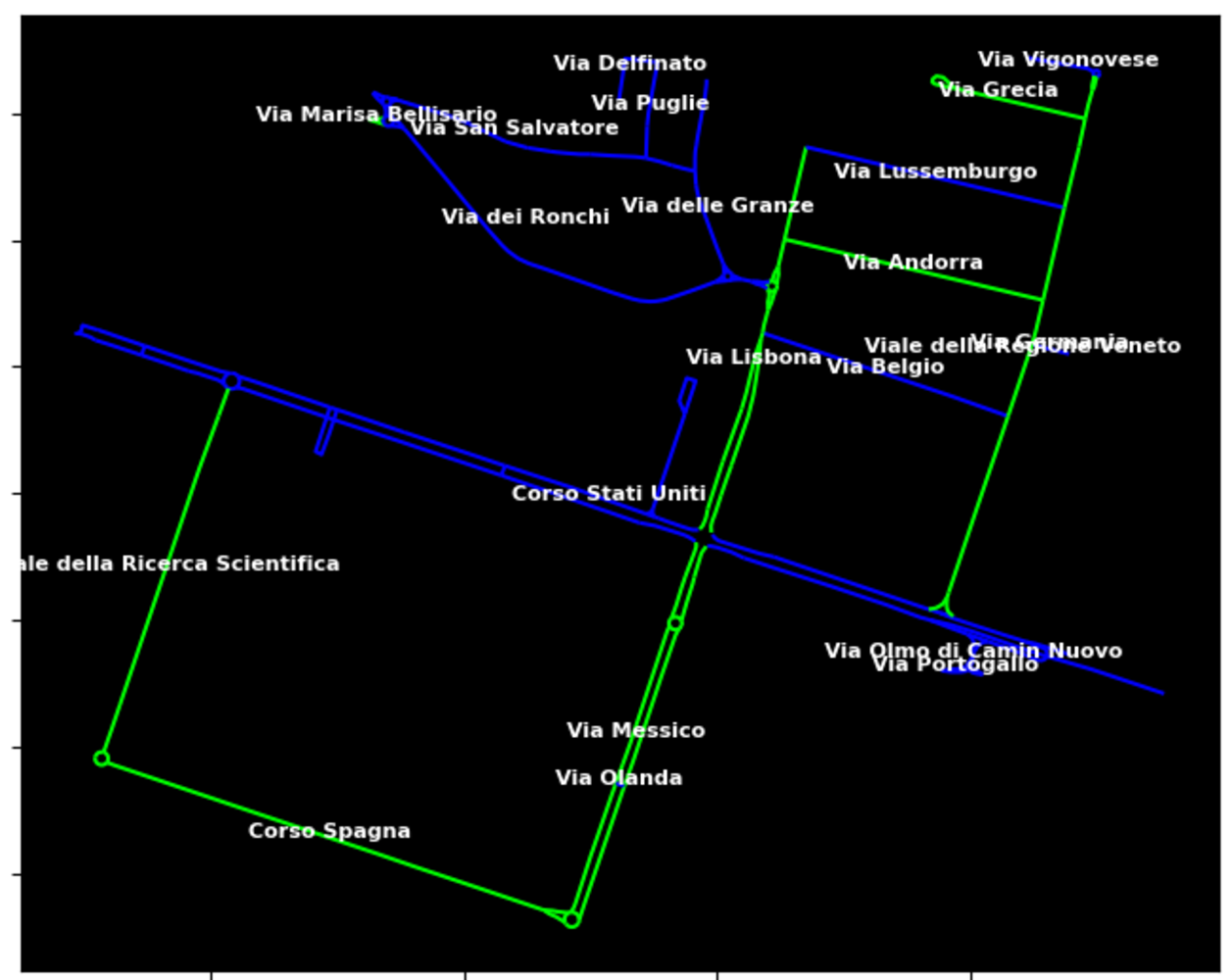}
	\end{minipage}
	\caption{The %MDPI: Please confirm if need add explanation for the blue lines.
		segments highlighted in green indicate the best selection according to the metrics $\mathrm{rank}[\mathcal{W}_{n}]$ (\textbf{left}) and $\mathcal{K}[\mathcal{W}_{n}]$ (\textbf{right}) of $p^{\star}=8$ roads among $p=22$ possible roads in the industrial zone of Padua, Italy. Whereas, blue segments indicate roads that are not selected. More on these simulations at \url{https://thesis.unipd.it/handle/20.500.12608/74384} (accessed on 21 October 2024).} %MDPI: Please add the access date (Format: Date Month Year). e.g., (accessed on 1 January 2020). Please note that the access date should be before the paper received date 14 January 2025.
	\label{fig:sss}
\end{figure}

In light of this premise, the~optimal sensor selection therefore consists of selecting a subset of $p^{\star}$, possibly redundant, sensors from a larger set of $p \gg p^{\star}$ potential sensors in order to preserve the structural observability property of the entire traffic network. 
An illustrative example of this approach to street sensor selection, whose problem formalization and solution implementation are yielded in the next paragraphs, is depicted in Figure~\ref{fig:sss}. {Detailed discussions on experimental results and case analyses related to the topic can be also found in~\cite{varotto2021probabilistic}.}

\subsection{Formalization of the~Problem}\label{ssec:problem_statement}

In this section, the~rigorous mathematical formulation of the street sensor selection problem is provided.
We consider an LTI system $\Sigma$ of the form \eqref{eq:state-space} 
describing the traffic flow in a given road network. 
Hereafter, the~attention is focused on determining the placement of $p^{\star}$ dedicated sensors, each of them measuring a single output variable, that is, the traffic flow of a single road within an urban network of $p$ roads. 

Given the dynamic relation in~\eqref{eq:state-space}, we can interpret the system matrices $C$ and $B$ as follows: each row of the matrix $C$ represents a sensor of the system, whereas each column of the matrix $B$ represents an actuator of the system. One has
\begin{equation}\label{eq:CB}
	C = \begin{bmatrix}
		c_1^\top & c_2^\top & \cdots & c_p^\top
	\end{bmatrix}^\top,\quad
	B = \begin{bmatrix}
		b_1 & b_2 & \cdots & b_m
	\end{bmatrix},
\end{equation}
where $c_j \in \mathbb{R}^{1\times n}$, $j = 1,2, \ldots, p$ represents a sensor, whereas $b_i \in \mathbb{R}^n$, $i = 1,2, \ldots, m$, represents an~actuator.

As we want to select only $p^{\star} \leq  p$ out of the $p$ sensors available for use at each time instant, we seek a selection matrix
\begin{equation}\label{eq:selection_matrix}
	S(\mathcal{Q}) = \begin{bmatrix}
		\mathrm{e}_{q_1} &
		\mathrm{e}_{q_2} &
		\cdots &
		\mathrm{e}_{q_{p^\star}}
	\end{bmatrix}^{\top},
\end{equation}
with $\mathcal{Q} = \{q_1,q_2, \ldots, q_{p^\star}\} \subseteq \mathcal{S} := \{1,2, \ldots, p\}$ that will choose which rows of the sensing matrix $C$ will be used, so that the new sensing matrix is $C_{\mathcal{Q}}: = S(\mathcal{Q}) C$. 
%This problem has been studied extensively in the literature and various algorithms to tackle it efficiently have been proposed. 
In this review, we consider the sensor selection problem under a restrictive setup in which only the output provided by the selected sensors at the instant $t \in \mathbb{N}$ is available for measurement. As~a minimum requirement, one may impose the following assumption, which implies knowing that at least one set of sensors renders the system observable. {In other words, such an assumption guarantees the feasibility of the street sensor selection problem}.

\begin{Asm}\label{asm:feasibility}
	For some $\tilde{p}<p$, there exists a known selection $\tilde{\mathcal{Q}} = \{\tilde{q}_1,\tilde{q}_2, \ldots, \tilde{q}_{\tilde{p}}\} \subset \mathcal{S}$, such that $(A, S(\tilde{\mathcal{Q}})C)$ is observable.
\end{Asm}

To describe the road quantities selected by the configuration $\mathcal{Q}$ in a more convenient fashion, we redefine $\Sigma$ via the following relation:
\begin{equation}\label{eq:SigmaQ}
	\Sigma^{\mathcal{Q}}: \begin{cases}
		x(t+1) = Ax(t) + Bu(t), \quad x(0)=x_0, \\
		\hat{y}_s(t) = S(\mathcal{Q}) (y(t) -Du(t)) = C_{\mathcal{Q}} x(t), \\
		\tilde{y}_s(t) = S(\tilde{\mathcal{Q}}) (y(t) -Du(t)) = C_{\tilde{\mathcal{Q}}} x(t)
	\end{cases} 
\end{equation}
so that the observability Gramian of $\Sigma^{\mathcal{Q}}$ associated to the output $\hat{y}_s(t)$ is given by
\begin{equation}\label{eq:Gramian_opt}
	\mathcal{W}^\mathcal{Q}_\tau = \sum_{k=0}^{\tau-1} (A^\top)^kC_{\mathcal{Q}}^\top C_{\mathcal{Q}} A^k.
\end{equation}

To simplify the notation, let us set $\mathcal{W}_{n}:=\mathcal{W}_{n}^{\mathcal{S}}$. It is worth noting that, whenever a configuration $\mathcal{Q}$ is selected satisfying the relation
\begin{equation}\label{eq:rankWWQ}
	\mathrm{rank}[\mathcal{W}_n] = \mathrm{rank}\left[\mathcal{W}^\mathcal{Q}_n\right] = n
\end{equation} 
then the following statements are~equivalent:
\begin{itemize}
	\item The couple $(A,C_{\mathcal{Q}})$ is observable;
	\item The couple $(A,C)$ is observable.
\end{itemize}
In other words, the~validity of relation \eqref{eq:rankWWQ} implies that the underlying system is observable with respect  to both the outputs $y(t)$ and $\hat{y}_s(t)$. A~similar result can be demonstrated with regard to detectability. More precisely, if~for a certain configuration $\mathcal{Q}$, the couple $(A,C_{\mathcal{Q}})$ is detectable, then the couple $(A,C)$ is detectable.

Following this rationale, %and considering the conclusion in Proposition~\ref{prop:detectability_of_selection}, 
the street sensor selection problem can be thus rigorously formalized as stated in the next~lines.

\begin{Problem}\label{pr:ss} 
	Given a fixed number of sensors $p^{\star}$ such that $1 \leq p^{\star} < p$, select a configuration $\mathcal{Q}^{\star}$ among all the configurations $\mathcal{Q}$ of $p^{\star}$ sensors for~which the following conditions are true: %EE: Please check that intended meaning was retained
	\begin{enumerate}
		\item[(i)] $\mathcal{Q}^{\star}$ is the result of a polynomial-time optimization having as optimality criterion the maximization of an observability-Gramian-based metric $f$ 
		computed on the observability Gramian~\eqref{eq:Gramian_opt}, such as those reported in Table~\ref{tab:gramian_metrics};
		\item[(ii)] the resulting couple $(A,S(\mathcal{Q}^{\star})C)$ is, possibly, detectable.
	\end{enumerate}
\end{Problem}

\subsection{Basic Algorithmic~Solutions} \label{sec:algorithms}

In the following lines, we shall illustrate a couple of elementary algorithmic approaches to solve Problem~\ref{pr:ss}. More formally, given a metric $f$ operating on the observability Gramian of the LTI system~\eqref{eq:SigmaQ}, we seek an algorithmic solution of
\begin{equation}\label{eq:problem0}
	\begin{split}
		\mathcal{P}_0:~&\text{find $\mathcal{Q}^{\star}$ from \eqref{eq: generalOptProblem}} \\
		&\text{subject to:}\  (A,C_{\mathcal{Q}^{\star}}) \text{ is detectable,}\\
		& \quad\quad\quad\quad\quad\text{with $C_{\mathcal{Q}^{\star}} = S(\mathcal{Q}^{\star})C$.} 
	\end{split} 
\end{equation}

%The combinatorial optimization problem of sensor selection in \eqref{eq: generalOptProblem} can be solved by exhaustive search. However, testing all combinations becomes intractable in case of large networks~\cite{Intractability}. For this reason, two heuristic strategies are proposed to address the problem in \eqref{eq: generalOptProblem}; the final result is the choice of an appropriate set of sensors $\mathcal{S}^* \subseteq \mathcal{S}$ such that $\vert \mathcal{S}^* \vert = p$. It is worth to remark that, if we denote as $ \mathcal{S}^{opt}$ the global optimum obtained by exhaustive search, it is not always guaranteed that $\mathcal{S}^* = \mathcal{S}^{opt}$; in fact, heuristic optimization procedures may find a local minimum in case of non-convexity problem. With reference to \mbox{Sec.~\ref{ssec: LTI}} and \eqref{eq: generalOptProblem}, the maximization of $\mathrm{rank}(\mathbf{W})$ and $\mathrm{tr}(\mathbf{W})$, or the minimization of $\mathcal{K}(\mathbf{W})$ are considered.

\begin{algorithm}[th!]
	\caption{Exhaustive-search-based selection~algorithm}\label{alg:exhaustive_search}
	\begin{algorithmic}[1]
		\REQUIRE system matrices, observability-based metric $f$, set of sensors $\mathcal{S}$, with~$|\mathcal{S}|=p$, number of sensors $p^{\star}\leq p$ to be selected
		\ENSURE set of selected sensors $\mathcal{Q}^{\star}\subseteq \mathcal{S}$, with~$|\mathcal{Q}^{\star}|=p^{\star}$ and $(A,C_{\mathcal{Q}^{\star}})$ detectable, the~flag $detectable=1$; or $\mathcal{Q}^{\star}$ attaining the highest score according to the given metric $f$ over all the trials and the flag $detectable=0$, if~none of the trials have led to a detectable couple $(A,C_{\mathcal{Q}^{\star}})$, that occurs if and only if the problem is infeasible
		\STATE $\mathcal{Q}:=\{q_1,q_2,\ldots,q_{p^{\star}}\} \leftarrow \{0,0,\ldots,0\}$
		\STATE $(\mathcal{Q}^{\star},score,detectable) \leftarrow (\varnothing,-\infty,0)$
		\STATE $(\mathcal{Q}^0,score^0) \leftarrow (\varnothing,-\infty)$
		%\STATE $count \leftarrow 2^{p}-2^{p-p^{\star}}$
		\STATE $cw \leftarrow [\underset{p^{\star}}{\underbrace{1 \cdots 1}} ~ \underset{p-p^{\star}}{\underbrace{0 \cdots 0}}]$
		\FOR{each codeword $cw$ such that $\sum\nolimits_{j=1}^{p} cw_{j} = p^{\star}$}
		% $count = 2^{p}-2^{p-p^{\star}},2^{p}-2^{p-p^{\star}}-1,\ldots, 2^{p^{\star}}-1$
		%\STATE $cw \leftarrow \mathrm{dec2bin}(count)$
		%\IF{$\sum\nolimits_{j=1}^{p} cw_{j} = p^{\star}$}
		\STATE $k \leftarrow 1$
		\FOR{$i=1,2,\ldots,p$}
		\IF{$cw_{i} = 1$}
		\STATE $q_k \leftarrow i$
		\STATE $k \leftarrow k+1$
		\ENDIF
		\ENDFOR
		\IF{$f(\mathcal{Q})>score$ and $(A,C_\mathcal{Q})$ is detectable}
		\STATE $(\mathcal{Q}^{\star},score,detectable) \leftarrow (\mathcal{Q},f(\mathcal{Q}),1)$
		\ENDIF
		\IF{$f(\mathcal{Q})>score^0$}
		\STATE $(Q^0,score^0) \leftarrow (Q,f(Q))$
		\ENDIF
		%\ENDIF
		%\STATE decrease $cw$
		\ENDFOR
		\IF{not $detectable$}
		\STATE $\mathcal{Q}^{\star} \leftarrow \mathcal{Q}^0$
		\ENDIF
		\RETURN $(\mathcal{Q}^{\star},detectable)$
	\end{algorithmic}
\end{algorithm}

\subsubsection{Exhaustive Search~Approach}

A potential approach to solve the problem of street sensor selection expressed by $\mathcal{P}_0$ is to examine all possible combinations of sensors and select the configuration that maximizes one of the metrics among those examined in Section~\ref{ssec:obs_metrics}. To~this purpose, Algorithm~\ref{alg:exhaustive_search} allows to find a configuration $\mathcal{Q}^{\star}$ that also represents the global optimum (satisfying the detectability requirement) attained by maximization in~\eqref{eq: generalOptProblem}. Such a property is guaranteed since the value ($score$) of a chosen observability-based metric $f$ is evaluated once for each possible configuration {(in particular, each configuration $\mathcal{Q}$, with~$|\mathcal{Q}|=p^{\star}$, is associated to a binary codeword $cw = \begin{bmatrix}
		cw_{1} & cw_{2} & \cdots & cw_{p}
	\end{bmatrix}$ composed of $p^{\star}$ ones and $p-p^{\star}$ zeros. %Also, concerning Algorithm~\ref{alg:exhaustive_search}, the~subroutine function $\mathrm{dec2bin}$ takes as input an integer number and returns its conversion into binary array. 
	Notice that %both the use of $\mathrm{dec2bin}$ 
	the check $\sum\nolimits_{j=1}^{p} cw_{j} = p^{\star}$ (line 5 in Algorithm~\ref{alg:exhaustive_search}) can be actually avoided by properly shifting ones and zeros in $cw$, thus yielding a more efficient implementation for this kind of search)  
	%, subroutine functions $\mathrm{binary\_array}$ and $\mathrm{to\_base10}$ are respectively employed to convert a decimal integer into a binary array of length $p$ and vice-versa.
} $\mathcal{Q}:=\{q_1,\ldots,q_{p^{\star}}\}$ containing $p^{\star}$ active sensors and maximized in a brute-force fashion. Consequently, we relabel the solution $\mathcal{Q}^{\star}$ obtained from the exhaustive search approach as $\mathcal{Q}^{opt}$. 

Although this procedure guarantees a totally predictable behavior from a mathematical point of view and exactly solves the original problem, it could require a great computational power in the case of large urban networks, scaling exponentially as the number of selected sensors $p^{\star}$ increases~\cite{Intractability}. As~already mentioned above, this drawback is due to the intrinsic combinatorial complexity related to the evaluation of all possible sensor configurations. For~this reason, further strategies requiring a lower computational burden can be employed, such as the random approach (see next subsection). For~this kind of strategy, the~final result obtained through such approaches is the choice of an appropriate set of sensors $\mathcal{Q}^\star \subseteq \mathcal{S}$ for which it is not always guaranteed that $\mathcal{Q}^\star = \mathcal{Q}^{opt}$; in fact, heuristic-based optimization procedures may generally find a local minimum whenever the given problem is nonconvex. %With reference to \mbox{Sec.~\ref{ssec: LTI}} and \eqref{eq: generalOptProblem}, the~maximization of $\mathrm{rank}(\mathbf{W})$ and $\mathrm{tr}(\mathbf{W})$, or~the minimization of $\mathcal{K}(\mathbf{W})$ are~considered.

%\vspace{-12pt}

\begin{algorithm}[h!]
	\caption{Random selection~algorithm}\label{alg:random}
	\begin{algorithmic}[1]
		\REQUIRE system matrices, observability-based metric $f$, set of sensors $\mathcal{S}$, with~$|\mathcal{S}|=p$, number of sensors $p^{\star}\leq p$ to be selected, $\alpha>0$ such that the total number of trials is exactly given by $\lceil \alpha\binom{p}{p^{\star}} \rceil$
		\ENSURE set of selected sensors $\mathcal{Q}^{\star}\subseteq \mathcal{S}$, with~$|\mathcal{Q}^{\star}|=p^{\star}$ and $(A,C_{\mathcal{Q}^{\star}})$ detectable, the~flag $detectable=1$; or $\mathcal{Q}^{\star}$ attaining the highest score according to the given metric $f$ over all the trials and the flag $detectable=0$, if~none of the trials have led to a detectable couple $(A,C_{\mathcal{Q}^{\star}})$
		\STATE $\mathcal{Q}:=\{q_1,q_2,\ldots,q_{p^{\star}}\} \leftarrow \{0,0,\ldots,0\}$
		\STATE $(\mathcal{Q}^{\star},score,detectable) \leftarrow (\varnothing,-\infty,0)$
		\STATE $(\mathcal{Q}^0,score^0) \leftarrow (\varnothing,-\infty)$
		\FOR{$j=1,\ldots,\lceil \alpha\binom{p}{p^{\star}} \rceil$}
		\STATE $idx \leftarrow \begin{bmatrix}
			1 & 2 & \cdots & p
		\end{bmatrix}$
		\FOR{$k=1,2,\ldots,p^{\star}$}
		\STATE $i \leftarrow \mathrm{DiscreteUniform}(1,p-k+1) $
		\STATE $q_k \leftarrow idx_i$
		\STATE $idx_i \leftarrow idx_{p-k+1}$
		\ENDFOR
		\IF{$f(\mathcal{Q})>score$ and $(A,C_\mathcal{Q})$ is detectable}
		\STATE $(\mathcal{Q}^{\star},score,detectable) \leftarrow (\mathcal{Q},f(\mathcal{Q}),1)$
		\ENDIF
		\IF{$f(\mathcal{Q})>score^0$}
		\STATE $(Q^0,score^0) \leftarrow (Q,f(Q))$
		\ENDIF
		\ENDFOR
		\IF{not $detectable$}
		\STATE $\mathcal{Q}^{\star} \leftarrow \mathcal{Q}^0$
		\ENDIF
		\RETURN $(\mathcal{Q}^{\star},detectable)$
	\end{algorithmic}
\end{algorithm}

\subsubsection{Random~Approach}\label{sez:approccio_proposto}

In this second approach, $p^{\star}$ sensors are randomly selected over $p$. Algorithm~\ref{alg:random} mainly consists of a loop of $\lceil \alpha \binom{p}{p^{\star}} \rceil$ iterations, where the parameter $\alpha>0$ is chosen. At~each iteration, the~selection indices $\{q_1,q_2,\ldots,q_{p^{\star}}\}$ are then assigned as the outcome of a uniform distribution, such that these can select $p^{\star}$ sensors with no need to repeat the execution (since exactly $p^{\star}$ random indices are generated at a time). The~set of selected sensors $\mathcal{Q}:=\{q_1,q_2,\ldots,q_{p^{\star}}\}$, whose values might be repeatedly assigned across the main iterations, is then compared with the best configuration $\mathcal{Q}^{\star}$ to find the highest $score$. Clearly, it is not ensured that $\mathcal{Q}^{\star}$ coincides with the optimal solution $\mathcal{Q}^{opt}$. And~this is not even guaranteed to occur as $\alpha \gg 1$, since any generation of the indices through the $\mathrm{DiscreteUniform}$ random variable is independent from the previous ones. Another drawback is the fact that no criterion can be used to break the main loop in this algorithm before its execution stops. Nonetheless, such a random approach has the advantage of searching significantly diverse sensor configurations at each step. Hence, Algorithm~\ref{alg:random} can be, in~principle, combined with any other approach not ensuring optimality (discussed in the sequel) to allow a strategic warm start or boost variables' space exploration in spite of~exploitation.

\subsection{State-of-the-Art Solutions to the Street Sensor Selection~Problem}

Within the framework of street sensor selection for traffic monitoring, the~description given in~\eqref{eq:state-space} comes in handy to describe, in~a simplified fashion and from a macroscopic point of view, an~urban network flow of vehicles. Frequently, each component of the state $x(t)$ is associated with the number of vehicles per second passing through a certain road connection. The~components of input $u(t)$ can be intended as the number of incoming vehicles with respect to the frontier nodes of the underlying network. The~state and input origin--destination matrices $(A,B)$ describe the state temporal update in the function of previous values $x(t)$ and the input $u(t)$. On the other hand, the~entries of $C$ are often binary and indicate the links on which sensors are placed. Usually, $D$ is set to zero while observing the output $y(t)$ selected through $C$. As~stated in Problem \ref{pr:ss}, the~matrix $C$ needs to be possibly found by optimizing one of the observability-Gramian-based metrics. The~particular choice of the latter key performance indicators is crucial, as~also demonstrated by the investigations reported in the next~lines.  

Problem \ref{pr:ss} has firstly been addressed in terms of observability in~\cite{georges1995use}, where the optimal location of sensors is viewed as the problem of maximizing the output energy generated by a given state. In~the dissertation~\cite{Hinsons}, the author formalizes the observability-based sensor placement problem by proposing a number of observability metrics to be optimized in order to select the locations of the~sensors.

Many variations of this problem have then become popular in the literature, especially those related to the optimal choice of street sensors to be deployed on road networks with the purpose of traffic flow monitoring. Nonetheless, the~main challenge arising while tackling Problem \ref{pr:ss} lies in its combinatorial nature, which is common across all its variants. Indeed, the~study in~\cite{rohloff2006approximating} shows that the problem of selecting a set of sensors of minimum cost used for the synthesis of a supervisory controller is related to a type of directed graph $st$-cut and can be converted into an integer programming problem. On~the other hand, ref. %MDPI: Newly added  information. Please confirm.
\cite{gagliardi2024joint} addresses the joint sensor selection and observer design problem for positive systems. The~proposed methodology leverages {Mixed-Integer} Semidefinite Programming techniques, providing a rigorous and systematic framework for optimizing the selection of sensors and the design of the corresponding optimal $L_1$ observer in a unified~manner.

The research in~\cite{joshi2008sensor} employs a convex optimization approach to simplify the combinatorial nature of sensor selection, combining convex relaxation with local optimization to efficiently obtain near-optimal solutions. The~concept of virtual variance, proposed in~\cite{VirtualVariances}, then transforms the sensor placement problem into a convex optimization task, reducing computational complexity compared to the original combinatorial formulation. In~addition, the~researchers in~\cite{NearOptimalGP} point out the drawbacks of entropy-based sensor placement approaches, which frequently result in suboptimal coverage at the boundaries of the region of interest. To~overcome this issue, they utilize Gaussian processes to model sensor measurements and maximize the mutual information between sensors, thereby achieving improved spatial coverage of~information. 

The authors of~\cite{summers2015submodularity} address this task by proposing different metrics based on the observability Gramian that can be maximized through a greedy approach exploiting submodularity with respect to  some of the aforementioned observability-Gramian-based metrics. This allows for solutions having an optimality gap upper-bounded by $36.8\%$, as~proven in~\cite{Li2009submodularity}. Similarly, the~work in~\cite{he2017vehicle} formulates the problem of camera placement in high-traffic urban areas as a submodular set function optimization problem, solved using a greedy heuristic with provable suboptimality bounds~\cite{Intractability}. Other heuristics are instead based on the simulated annealing strategy~\cite{gagliardi2023traffic}, which solves the problem under a predefined budget constraint on the maximum number of sensors to be used. 
The results presented in~\cite{varotto2019street,staroswiecki2004sensor} are also considered. The~first work proposes
an end-to-end solution to the street sensor selection problem
through road network modeling and observability
measures. On the other hand, the~second  work addresses the problem of fault
tolerant estimation and the design of fault tolerant sensor
networks, introducing the concepts of redundant and
minimal sensor sets and organizing them into an automaton, i.e.,~a control mechanism designed to automatically follow a sequence of operations. 
Extensively explored in the literature, further works have proposed various algorithms to maximize $\mathcal{H}_2$ norm-related metrics, capitalizing on the assumption that more powerful sensors yield stronger signals~\cite{Manohar2022Optimal,clark2020sensor,chen2011h2,summers2014optimal}. Additionally, sensor selection has been approached from different angles, not only including the optimization of Gramian-related metrics~\cite{summers2016convex,yamada2023efficient,bopardikar2017sensor,siami2020deterministic}, but~also linear--quadratic regulation~\cite{dhingra2014admm,zare2019proximal,tzoumas2020lqg} and security concerns~\cite{milovsevic2020actuator,pirani2021game}.

However, many existing methodologies presuppose complete knowledge of the system model, which may be restrictive in some scenarios. 
Addressing the challenge of unknown system dynamics, adaptive control techniques such as model-free optimal control~\cite{kiumarsi2017optimal,jiang2014robust,modares2014optimal} and adaptive backstepping~\cite{krstic1995nonlinear} have been proposed. Recent advancements in learning algorithms have further facilitated sensors and actuators selection in dynamic environments~\cite{vafaee2024real,ye2022online,fotiadis2023data,fotiadis2024learning}. Notably, real-time learning-based approaches have emerged to dynamically decide sensor utilization based on evolving system behavior~\cite{vafaee2024real}. Yet, limitations persist, particularly in scenarios where only output data are accessible, rendering sensor selection considerably more~complex.

\indent Lastly, we report on new trends in sensor selection for systems with unknown characteristics, focusing on input--output data. Despite the inherent challenges, leveraging techniques from reinforcement learning~\cite{lewis2010reinforcement}, a~cost function can be expressed solely based on input--output data. These data-driven approaches enable effective street sensor selection even in scenarios where system matrices are unknown, offering a promising avenue for addressing the constraints of the framework under~analysis.

%\end{comment}

%These techniques will be surveyed in the next subsection.

%%%%%%%%%%%%%%%%%%%%%%%
%%%%%%%%%%%%%%%%%%%%%%%
%%%%%%%%%%%%%%%%%%%%%%%
%%%%%%%%%%%%%%%%%%%%%%%
\section{Discussion}\label{sec:discussion}

In this study, the~efficient sensor selection for traffic flow monitoring has been addressed. So far, we have provided an overview of the most popular and powerful model-based techniques leveraging the notion of road network observability.
Nevertheless, a~few limitations should be pointed out with respect to  these solutions. Traditionally, traffic modeling relies on the identification of system matrices to capture the underlying dynamics accurately. However, this approach is often hindered by the intricate and nonlinear nature of traffic systems, leading to inaccuracies or inadequacies in model representation. {In addition, the~full reliance on state-space models is also related to another major drawback: since the number of possible placement configurations $\binom{p}{p^{\star}}$ increases super-exponentially ({by leveraging Stirling's approximation~\cite{robbins1955remark}, it can be proven that the quantity $\binom{p}{p^{\star}}$ grows no faster than $O((2p/\exp(1))^{p^{\star}}) \simeq O((0.736p)^{p^{\star}})$, as~$p=2p^{\star} \rightarrow +\infty$ (the worst case scenario occurs for $p^{\star} = p/2$)}) as the numbers of available and deployed sensors $p,p^{\star}$ grow, then the NP-hardness for brute-force optimal algorithmic solutions, such as Algorithms \ref{alg:exhaustive_search} and  \ref{alg:random}, is established.} Data-driven methodologies, such as reinforcement learning and machine learning techniques, are thus envisaged to circumvent the need for explicitly identifying system matrices. An increasing adoption of data-driven approaches grounded in the observability of networked systems is foreseen for sensor selection in traffic modeling. This shift represents a significant advancement in addressing the complexities inherent in traffic flow dynamics and, ultimately, in~establishing innovative and explainable methods for devising efficient ITSs.
Indeed, besides~assigning a set of sensors for traffic flow monitoring, such data-driven methods also allow real-time scheduling in uncertain scenarios (see, e.g.,~\cite{gupta2006stochastic}) for the already deployed WSN and the employment of additional mobile sensors. In~online sensor scheduling, the~selection of sensors is indeed made dynamically over time, as~opposed to being predetermined in advance. At~each time step, the~sensor set is evaluated based on the current state of the system, and~a decision is made about which sensors to keep or discard.
As highlighted in~\cite{vafaee2024real}, the~key characteristic of online sensor scheduling is its causality, meaning the decisions made at each time step are based only on the state knowledge up to the current time, without~knowing the future. This makes the problem challenging because the scheduling decisions must account for the uncertainty and changing dynamics of the system without complete future information.
In fact, the~objective of online sensor scheduling is to minimize the use of sensors while maintaining the observability of the system, as~close as possible to the fully-sensed~dynamics.

The adoption of data-driven sensor selection methodologies {rooted in network observability would} not only enhance the efficacy of sensor deployment and mesoscale or microscale traffic modeling but also catalyze advancements in communication technologies, thus fostering the IoV paradigm. In~order to facilitate the transmission of large datasets required for rescheduling, improvements in communication infrastructure are imperative. This includes the development of high-speed, reliable communication networks capable of handling the influx of data from distributed sensors in real time based on 5G, 6G, and MEC technologies.
Moreover, given the critical importance of energy efficiency in IoT-oriented WSNs, additional efforts to minimize energy consumption during data transmission are paramount. This involves the redesign and implementation of energy-efficient communication protocols and algorithms to optimize data transmission while conserving energy~resources.\\
\indent We believe that the interdisciplinary nature of data-driven sensor selection approaches resonates with stakeholders across transportation, IoT, and~control systems domains. By~addressing the challenges of traffic modeling and monitoring along with sensor selection and scheduling through innovative data-driven techniques (such as data-driven predictive control~\cite{piga2017direct,dorfler2022bridging,BreschiFabrisFormentin2023IFAC,BreschiChiusoFabris2023CDC,ChiusoFabris2025}), these novel methodologies are envisaged for reaching more efficient and adaptive ITSs, with~broader implications for urban planning, traffic management, and~infrastructure~development.

\section{Concluding Remarks and Future~Directions}\label{sec:conclusions}

This paper reviews the main model-based techniques leveraging network observability to perform efficient sensor selection for traffic flow monitoring. It is highlighted how modeling and monitoring urban networks constitute an interdependent challenging task. To~this aim, advantages and current limitations with respect to  the state-of-the-art approaches are discussed, promoting the use of data-driven sensor selection strategies {inspired by network observability} together with the advancement of cutting-edge 6G communication technologies towards the Interned-of-Vehicles paradigm. Tackling the difficulties associated with traffic modeling, monitoring, sensor selection, and~scheduling through cutting-edge data-driven methodologies holds the potential to realize more efficient and adaptive intelligent transportation systems. These innovative approaches carry wide-ranging implications for urban planning, traffic management strategies, and~the development of transportation~infrastructure.

\indent Future work linked to these topics will consider the investigation of new observability-based metrics for time-varying or nonlinear systems, {as well as the research devoted to the comparison of the existing metrics}, and~the %development 
{innovation} of data-driven strategies to overcome stringent assumptions related to the adoption of rigid model~structures.

\section*{Funding}
This study was carried %MDPI: Information regarding the funder and the funding number should be provided. Please check the accuracy of funding data and any other information carefully. Any updates after publication should be carefully considered.
	out within the MOST (the Italian National Center for Sustainable Mobility) and received funding from NextGenerationEU (Italian PNRR---CN00000023---D.D. 1033 17/06/2022---CUP C93C22002750006).

\section*{Acknowledgments}
%This work has been partly supported by the Grant ``Design Of Cooperative Energy-aware Aerial plaTforms for remote and contact-aware operations'' (DOCEAT) funded by the Italian MUR, n. 2020RTWES4, CUP n. E63C22000410001.

The authors would like to express their gratitude to Aynur \c{S}en for her valuable assistance with coding and producing the images presented in Figure~\ref{fig:sss}.

\bibliography{biblio}

\end{document}